\documentstyle[aps,preprint]{revtex}
 \tightenlines
\begin{document}
\draft
\title{Inverse Avalanches On Abelian Sandpiles}
\author{H. F. Chau$^{1,2}$}
\address{
 ~$^1$School of Natural Sciences, Institute for Advanced Study, Olden Lane,\\
 Princeton, NJ 08540, U.S.A.\\
 ~$^2$Department of Physics, University of Illinois at Urbana-Champaign,\\
 1110 West Green Street, Urbana, IL 61801-3080, U.S.A.
}
\date{\today}
\preprint{IASSNS-HEP-94/77}
\maketitle
\mediumtext
\begin{abstract}
 A simple and computationally efficient way of finding inverse avalanches
 for Abelian sandpiles, called the inverse particle addition operator, is
 presented. In addition, the method is shown to be
 optimal in the sense that it requires the minimum amount of computation
 among methods of the same kind. The method is also conceptually nice because
 avalanche and inverse avalanche are placed in the same footing.
\end{abstract}
\medskip
\pacs{PACS numbers: 05.40.+j, 05.60.+w, 64.60.Ht}
\narrowtext
 The Abelian Sandpile Model (ASM), whose mathematical structure is first
 studied extensively by Dhar \cite{ASM}, is one of the few class of models of
 self-organized criticality in which a lot of interesting physical properties
 can be found analytically. The model consists of a finite number of sites
 labeled by an index set $I$. For each site $i\in I$, we assign an integer
 $h_i$ called the local height to it. Whenever the local height of a site
 exceeds a threshold (which is fixed to 0 for simplicity), the site is called
 unstable and it will transport some of its local heights (or sometimes called
 particles at that site) to the other sites in
 the next timestep by
\begin{equation}
 h_j \longrightarrow h_j - \Delta_{ij} \hspace{0.3in} \mbox{whenever~} h_i > 0
 \mbox{.}
\end{equation}
 $\Delta$ is called the toppling matrix whose elements satisfies:
\begin{mathletters}
\begin{equation}
 \Delta_{ii} > 0 \hspace{0.25in} \forall\ i\in I \mbox{,} \label{E:Delta_Dia}
\end{equation}
\begin{equation}
 \Delta_{ij} \leq 0 \hspace{0.25in} \forall\ i\ne j \mbox{,}
 \label{E:Delta_OffDia}
\end{equation}
 and
\begin{equation}
 \sum_{j\in I} \Delta_{ij} \geq 0 \hspace{0.25in} \forall\ i\in I \mbox{.}
\end{equation}
\end{mathletters}
\par\noindent
 Toppling is repeated until all sites become stable again. The whole
 process of toppling is collectively known as an avalanche. The system is
 driven by adding unit amount of particles onto the sites randomly and
 uniformly after the system regains its stability.
\par
 A system configuration, stable or not, can be regarded as a point in the space
 ${\cal Z}^N$ where $N$ is the total number of sites in the system. Both
 the addition of a particle and the toppling of particles in a site can be
 regarded as a translation in ${\cal Z}^N$ \cite{ASM,GASM,FigureCut}. The
 process of adding a particle to the site $i$ together with the subsequent
 toppling it triggers (if any) can be viewed as a map between the set of all
 stable system configurations, and is denoted by ${\bf a}_i$ \cite{ASM,ASM2}.
\par
 Based on the observation that the final stable state is independent of the
 order of toppling in different sites, Dhar shows that ${\bf a}_i \circ
 {\bf a}_j (\alpha) = {\bf a}_j \circ {\bf a}_i (\alpha)$ for any stable system
 configuration $\alpha$ \cite{ASM}. This is why we call the model ``Abelian''.
 Using this commutative property, the total number of recurrence states in
 the model is shown to be $\det\Delta$ \cite{ASM}, and we denote the set of
 all recurrence states by $\Omega$.
\par
 Generalization of the ASM, known as the Generalized Abelian Model (GASM), has
 been made recently by Chau and Cheng \cite{GASM}. In their model, the local
 heights $h_i$ and the elements in the toppling matrix $\Delta$ are real
numbers
 instead of integers. Also, $\Delta$ satisfies only Eqs.~(\ref{E:Delta_Dia})
 and~(\ref{E:Delta_OffDia}). Arbitrary amount of particles are allowed to add
to
 the system in possibly different locations all at the same time. Moreover,
some
 special kind of configuration dependent triggering thresholds are used to
 determine the local stability of the pile (see ref.~\cite{GASM} for details).
 In spite of the large differences between ASM and GASM, similar commutative
 properties between particle addition operators are found for recurrence system
 configurations.
\par
 In both the ASM and the GASM, one can prove that for any pair of
 $\alpha\in\Omega$
 and ${\bf a}_i$, there exists a unique system configuration $\beta\in\Omega$
 such that ${\bf a}_i (\beta ) = \alpha$ \cite{ASM,ASM2}. While the avalanche
 problem (i.e. the problem of finding $\alpha$ given $\beta$ and ${\bf a}_i$)
 is straight forward and can be done very quickly in computer; the inverse
 avalanche problem (i.e. the problem of finding $\beta$ given $\alpha$ and
 ${\bf a}_i$) is much more difficult. $\beta$ cannot be found, in general, by
 simply removing particle(s) from site(s) because particle removal operation
 can map a state out of the eventual phase space $\Omega$. Furthermore, the
 relationship ${\bf a}_i^{-1} = {\bf a}_i^{\det\Delta -1}$ does not work very
 well for two reasons. First, $\det\Delta$ is in general a huge number making
 the method computationally impractical. Second, in the event of GASM, $\det
 \Delta$ may not be an integer and hence ${\bf a}_i^{\det\Delta}$ is not
 well defined.
\par
 The first computationally feasible method of finding inverse avalanches is
 proposed recently by Dhar and Manna using the so-called inverse avalanche
 operator by means of the burning algorithm
 \cite{Inverse_Aval}. However, the method works only on ASM with a symmetric
 toppling matrix. In this letter, we introduce a simple method to
 tackle the inverse avalanche problem for both ASM and GASM. To each particle
 addition operator, we find a nice and simple way to associate its inverse to
 another particle addition operator, called the inverse particle addition
 operator.
 Then we prove that the method is computationally optimal in the sense that it
 requires the least number of topplings among all the possible methods using
 the idea of inverse particle addition operators.
\par
 For simplicity, we concentrate only on the case of Abelian Sandpiles in the
 discussions below. However, all arguments, after slight modifications, work
 equally well on GASM. We represent a system configuration $\alpha$, stable or
 not, by a column vector of length $N$. In particular, the marginally stable
 state of the system (i.e. the one to which avalanche is triggered whenever
 particles are added to any one of the sites) is $\vec{0}\equiv (0,0,\ldots
 ,0)$.
\par
 Given a particle addition operator ${\bf a}$, we consider
\begin{equation}
 \gamma \equiv \left(\gamma_j\right)_{j\in I} \equiv {\bf a} (\vec{0}) = \left(
 a_j - \mbox{$\sum_k$} n_k \Delta_{kj} \right)_{j\in I} \in \Omega
 \label{E:Def_Gamma}
\end{equation}
 where $a_i\geq 0$ is the number of particles added to site $i$, and $n_i\in
 {\cal N}$ for all $i\in I$ is the total number of toppling in site $i$
 triggered after the particles are added \cite{FigureCut}. For example, $a_j =
 \delta_{ij}$ for the operation of adding a single particle to site $i$
 together with the subsequent toppling induced ${\bf a}_i$. The above
 definition works equally well when more than one particle is introduced to the
 system each time, and when they are introduced to different sites.
\par
 Consider the operation of adding $-\gamma_i$ particles to site $i$ for all
 $i\in I$ together with the subsequent toppling induced (if any). We denote
 this operation by $\tilde{\bf a}$. This is a well defined operation sending
 system configurations from $\Omega$ to $\Omega$ because $\gamma_i\leq 0$ for
 all $i$. For any $\alpha\equiv\left( \alpha_i \right)_{i\in I} \in
 \Omega$,
\begin{equation}
 \tilde{\bf a} (\alpha ) = \left( \alpha_j - a_j + \mbox{$\sum_k$} {n_k}'
 \Delta_{kj} \right)_{j\in I} \in \Omega
\end{equation}
 for some ${n_k}'\in {\cal Z}$. Moreover,
\begin{equation}
 {\bf a}\circ\tilde{\bf a} (\alpha ) = \left( \alpha_j + \mbox{$\sum_k$}
 {n_k}'' \Delta_{kj} \right)_{j\in I} \in \Omega
\end{equation}
 for some ${n_k}''\in {\cal Z}$. But by the remark between corollary~1 and
 corollary~2 in ref.~\cite{GASM}, we
 conclude that ${n_k}'' = 0$ for all $k\in I$ and hence  ${\bf a}\circ
 \tilde{\bf a} (\alpha) = \alpha$ for all $\alpha\in\Omega$. Since the particle
 addition operators commute with each other, we can also conclude that
 $\tilde{\bf a}\circ{\bf a} (\alpha ) = \alpha$. As a result, $\tilde{\bf a}
 \equiv {\bf a}^{-1}$ is the inverse particle addition operator corresponding
 to the
 particle addition operator ${\bf a}$.
\par
 We proceed to show that the above way of finding inverse avalanches
 is computationally optimal in the sense that the total number of toppling
 involve in the calculation is minimum among all the inverse particle addition
 operators (such as ${\bf a}^{\det\Delta -1}$).
\par
 Suppose ${\bf b}'$ is another inverse particle addition operator corresponding
 to the
 particle addition operator ${\bf a}$ consisting of adding $b_i (\geq 0)$
 particles to site $i$ for all $i\in I$ together with the subsequent toppling.
 Clearly, $b_j = \sum_k m_k \Delta_{kj} - a_j \geq 0$ for some $m_k \in
 {\cal N}$ \cite{GASM}. Consider the addition of $b_i$ particles into site $i$`
 for all
 $i$ at the same time to the system configuration $\gamma$. The resultant
 configuration is $\mu \equiv \left( \sum_k \left( m_k - n_k \right)
\Delta_{kj}
 \right)_{j\in I}$. Since ${\bf b}' (\gamma) = {\bf a}^{-1} (\gamma) =
 \vec{0}$, $\mu$ must either equal to $\vec{0}$ or it is an unstable
 configuration which will eventually topple to $\vec{0}$. In either case, we
 can conclude that $m_i - n_i \geq 0$ for all $i\in I$ and the equality holds
 if and only if $b_i = -\gamma_i$ for all $i\in I$.
\par
 For any $\alpha\in\Omega$, the introduction of $b_i$ particles to site $i$
 for all $i$ is equivalent to the addition of first $-\gamma_i$ particles to
 site $i$ for all $i$ and then $\sum_k ( m_k - n_k ) \Delta_{ki}$ particles
 into site $i$ for all $i$. So as compared to $\tilde{\bf a}$, ${\bf b}'$
 requires $\sum_i ( m_i - n_i )$ more toppling to find the inverse avalanche.
 Thus $\tilde{\bf a}$ is computationally optimal.
\par
 In summary, we have introduced a nice, simple and efficient way to find the
 inverse avalanche for both the Abelian Sandpile and the Generalized Abelian
 Sandpile by means of inverse particle addition operator. The method is
 conceptually nice because inverse particle addition operators
 are placed in the same footing as the particle addition operators.
\acknowledgments{This work is supported by DOE grant DE-FG02-90ER40542 and NSF
 grant AST-9315133.}

\end{document}